\documentclass[aps,final,twocolumn,showpacs,superscriptaddress]{revtex4}

\usepackage{graphicx}
\usepackage{amsmath}
\usepackage{amsfonts}
\usepackage{hyperref}
\usepackage{cool}

\newcommand{\bra}[1]   {\ensuremath{\langle#1|}}
\newcommand{\ket}[1]   {\ensuremath{|#1\rangle}}

\newcommand{\reddF}    {\ensuremath{(F'||d||F)}}
\newcommand{\fs}       {\ensuremath{\textrm{fs}}}
\newcommand{\hfs}      {\ensuremath{\textrm{hfs}}}
\newcommand{\e}        {\ensuremath{\vec \varepsilon}}

\newcommand{\es}       {\ensuremath{{\e}^{\,*}}}
\newcommand{\iee}      {\ensuremath{i(\e\wedge\es)}}

\newcommand{\ecez}     {\ensuremath{\e \cdot \vec e_Z}}
\newcommand{\eceB}     {\ensuremath{\e \cdot \vec e_{B_s}}}

\newcommand{\etes}     {\ensuremath{[\e \otimes \es]}}
\newcommand{\di}       {\ensuremath{\hat{\vec d}}}
\newcommand{\tg}       {\ensuremath{\tilde g}}
\newcommand{\beff}     {\ensuremath{{\vec b_\text{eff}}}}

\newcommand{\SYRTE}{\affiliation{LNE-SYRTE, Observatoire de Paris, CNRS, UPMC~; 61
avenue de l'Observatoire, 75014 Paris, France}}

\begin{document}

\title{Polarizabilities of the $^{87}$Sr Clock Transition}

\author{C. Shi} \SYRTE
\author{J.-L. Robyr} \SYRTE
\author{U. Eismann} \SYRTE
\author{M. Zawada} \SYRTE
\affiliation{Institute of Physics, Faculty of Physics, Astronomy and Informatics, Nicolaus Copernicus University, Grudzi\c{a}dzka 5, PL-87-100 Toru\'n, Poland}
\author{L. Lorini} \SYRTE
\affiliation{Istituto Nazionale di Ricerca Metrologica (INRIM), Strada delle Cacce
91, 10135 Torino, Italy}
\author{R. Le Targat} \SYRTE
\author{J. Lodewyck} \SYRTE

\begin{abstract}
	In this paper, we propose an in-depth review of the vector and tensor polarizabilities of the two energy levels of the $^{87}$Sr clock transition whose measurement was reported in~[P. G. Westergaard \emph{et al.}, Phys. Rev. Lett. \textbf{106}, 210801 (2011)]. We conduct a theoretical calculation that reproduces the measured coefficients. In addition, we detail the experimental conditions used for their measurement in two Sr optical lattice clocks, and exhibit the quadratic behaviour of the vector and tensor shifts with the depth of the trapping potential and evaluate their impact on the accuracy of the clock.
\end{abstract}

\pacs{06.30.Ft, 42.62.Fi, 37.10.Jk}
\maketitle

\section*{Introduction}

Optical lattice clocks are now the most stable frequency references~\cite{Hinkley13092013, 6468089, bloom_optical_2014}, and their accuracy is steadily improving towards $10^{-17}$ to $10^{-18}$~\cite{bloom_optical_2014, 1367-2630-16-7-073023, le_targat_experimental_2013, ushijima2015cryogenic}. In these clocks the systematic effects due to the atomic motion are cancelled by trapping  a large number of ultra-cold atoms (typically $10^4$) in the Lamb-Dicke regime, using an optical lattice formed by a standing wave laser beam. A specificity of these optical clocks is the strong light shift induced by the intense trapping light. At the so-called ``magic wavelength'', for which the scalar polarizabilities of the fundamental and excited clock states are identical, this light shift is largely suppressed~\cite{PhysRevLett.91.173005}. However, in order to achieve a high accuracy, the residual polarization-dependent and higher order light shifts have to be evaluated. In reference~\cite{westergaard2011lattice}, we reported exhaustive measurements of these light shifts with two $^{87}$Sr lattice clocks, and showed that they are compatible with an accuracy of $10^{-17}$ at a trapping potential of $100$ recoil energies. We reported the first experimental resolution of the vector-, tensor-, and hyper-polarizability and put an upper bound on higher order effects for $^{87}$Sr.

In this paper, we focus on the first-order electric dipole interaction to propose a theoretical study of the coefficient of the decomposition of its Hamiltonian in vector and tensor irreducible operators that can reproduce the experimental results reported in~\cite{westergaard2011lattice}. The first section introduces the irreducible operator formalism describing the atomic polarizability. In the second section, we make use of this decomposition to theoretically evaluate the vector and tensor polarizability coefficients. The last section details the measurement of these coefficients and show their agreement with the theoretical estimates. Finally, we describe the non-linear dependence of the vector and tensor shifts with the depth of the trapping potential. Furthermore, we report on precise measurements of the magic wavelength and its sensitivity that provides insight in the physical properties of electronic levels of $^{87}$Sr that are useful for characterizing the clock accuracy~\cite{PhysRevLett.109.263004}.

\section{Factorization of the polarizability operator}

The Hamiltonian that describes an atomic level $\ket{\phi} = \ket{nJFm}$ in the presence of the electromagnetic field of a trapping light and a static bias magnetic field $\vec B_s$ reads:
\begin{equation}
	\label{eq:H}
	\hat H = \hat H_Z + \hat H_\text{eff}.
\end{equation}
$\hat H_Z$ is the Hamiltonian representing the Zeeman interaction:
\begin{equation}
	\hat H_Z = \frac{g \mu_B}{\hbar} \hat {\vec F} \cdot \vec B_s,
\end{equation}
where $g$ is the Land\'e factor and $\mu_B$ the Bohr magneton. $\hat H_\text{eff}$ is the effective Hamiltonian representing the the off-resonant electric dipole interaction between the atom and the trapping light of complex amplitude $\mathcal{E}$ or power density $\mathcal{I} = \varepsilon_0c|\mathcal{E}|^2/2$ in second-order perturbation theory~\cite{PhysRevA.5.968}:
\begin{equation}
	\hat H_\text{eff} = - \frac 1 4 \hat \alpha |\mathcal{E}|^2,
\end{equation}
where the polarizability operator $\hat \alpha$ reads:
\begin{equation}
	\label{eq:pol}
	\hat \alpha = \frac{1}{\hbar} \sum_{\ket{\phi'}} \frac{\es \cdot \di\,\ket{\phi'}\bra{\phi'}\e \cdot \di}{\omega_0 - \omega_l} + \frac{\e \cdot \di\,\ket{\phi'}\bra{\phi'}\es \cdot \di}{\omega_0 + \omega_l},
\end{equation}
\begin{figure}
	\begin{center}
		\includegraphics[width=\columnwidth]{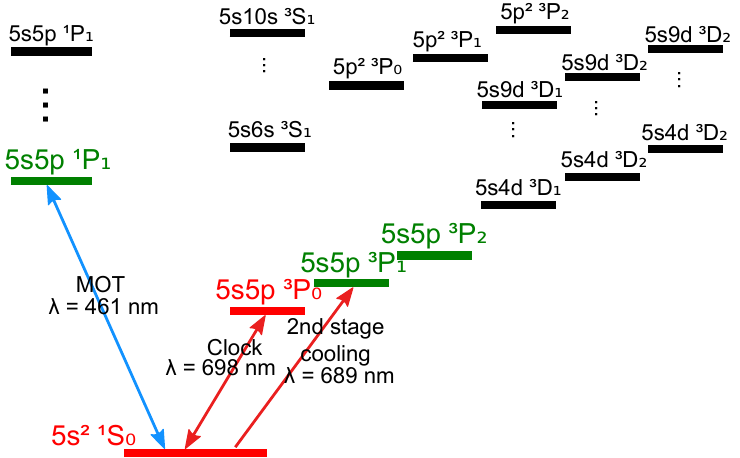}
	\end{center}
	\caption{\label{fig:levels}Energy levels for Sr. The two clocks states are represented in red. The $F = 9/2$ hyperfine  component of the three green states are involved in the vector and tensor polarizabilities of the excited clock state $5s5p\ {}^3P_0$, as explained in section~\ref{sec:theory}. The other states are considered to calculate the polarizability coefficients of the clock transition.}
\end{figure}
where the sum runs over all excited states $\ket{\phi'} = \ket{n'J'F'm'}$. $\di$ is the electric dipole operator, $\hbar\omega_0$ is the energy difference between $\ket{\phi'}$ and $\ket{\phi}$, and $\omega_l$ and $\e$ are the angular frequency and polarization of the trapping light, respectively. Because the two states involved in the clock transition ($5s^2\ {}^1S_0$ and $5s5p\ {}^3P_0$, as show in figure~\ref{fig:levels}) both have $J=0$, their respective polarizabilities do not \emph{a priori} depend on the magnetic sub-level $m$ considered. However, the hyperfine Hamiltonian describing odd Sr isotopes with a non-zero nuclear spin slightly breaks the rotational invariance and introduces a minute dependence of the polarizabilities of $^1S_0$ and $^3P_0$ on $m$ and $\e$. This dependence can be made analytically explicit by expanding the polarizability operator as the sum of three irreducible operators~\cite{romalis_zeeman_1999,ovsyannikov_polarisation_2006}. In the $(2F + 1)$-dimensional basis $\ket{m} \equiv \left\{\ket{nJFm}, -F \leq m \leq F\right\}$ of the hyperfine manifold of the Hilbert space describing either clock state, the matrix elements of this expansion read:
\begin{equation}
	\label{eq:polemfact}
	\bra{m_2}\hat \alpha\ket{m_1} = \sum_{j = 0}^{2} K_j \sum_{M = -j}^{j}(-1)^M C_{Fm_1j\,{-M}}^{Fm_2} \etes_{jM},
\end{equation}
where the first sum runs over the tensor rank $j$. In this expression, the dependence in $\e$ is exclusively contained in the tensor product $\etes_{jM}$, the dependence in $m$ in the Clebsch-Gordan coefficient $C_{Fm_1j\,{-M}}^{Fm_2}$ and the dependence in $\ket{\phi'}$ in the coefficients:
\begin{equation}
	K_j =  \frac{1}{\hbar} \sum_{\ket{n'J'F'}} k_j \left(\frac{1}{\omega_0 - \omega_l} + \frac{(-1)^j}{\omega_0 + \omega_l}\right),
\end{equation}
where $k_j$ is expressed using Wigner 6j-symbols and reduced dipole elements:
\begin{equation}
	\label{eq:kj}
	k_j = (-1)^{j-F-F'}\sqrt{\frac{2j+1}{2F+1}}\SixJSymbol{F,F,j}{1,1,F'}|\reddF|^2.
\end{equation}

The canonical scalar $\alpha_s$, vector $\alpha_v$ and tensor $\alpha_t$ polarizabilities are then defined by rescaling these coefficients:
\begin{equation}
\begin{split}
	\alpha_s = -\frac{1}{\sqrt{3}}\,K_0, \quad \alpha_v = \sqrt{\frac{2F}{F+1}}\,K_1,\\ \text {and} \quad \alpha_t = \sqrt{\frac{2F(2F-1)}{3(F+1)(2F+3)}}\,K_2.
\end{split}
\end{equation}
These three physical parameters are sufficient to completely explicit the matrix of the polarizability operator given by equation~(\ref{eq:polemfact}). The first term of this equation, for $j = 0$, is the scalar polarizability. It is the main, rotationally invariant contribution to the polarizability. The vector polarizability ($j = 1$) appears only when $F \geq 1/2$ and when the light has a non-linear polarization. It is equivalent to a fictitious magnetic field along the light wave vector. It is an odd function of $m$, and accordingly vanishes when the light shift is averaged over opposite values of $m$. The last term ($j = 2$), appearing if $F \geq 1$, is the tensor polarizability that results in a polarization dependent polarizability with an even dependence in $m$.

The following section is devoted to the theoretical calculation of the three polarizability coefficients for the two atomic levels involved in the $^{87}$Sr clock transition.

\section{Theoretical calculation of the polarizabilities}
\label{sec:theory}

In order to further calculate the polarizabilities, the $k_j$ coefficients can be related to the $\ket{\phi'} \rightarrow \ket{\phi}$ transition rate $\Gamma_{\phi\phi'}$, or to the  radiative lifetime $\tau$ of the excited state $\ket{\phi'}$ through:
\begin{equation}
	\frac{|\reddF|^2}{2F'+1} = B_\hfs \frac{3\pi\varepsilon_0\hbar c^3}{\omega_0^3} \Gamma_{\phi\phi'} = B_\hfs \frac{3\pi\varepsilon_0\hbar c^3}{\omega_0^3} B_\fs \frac{\zeta}{\tau},
\end{equation}
where the fine and hyperfine branching ratios read:
\begin{eqnarray}
	B_\fs  & = & (2J + 1)(2 L'+1){\SixJSymbol{J',1,J}{L,S,L'}}^2,\\
	B_\hfs & = & (2F+1)(2J'+1){\SixJSymbol{F',1,F}{J,I,J'}}^2
	\label{eq:Bhfs},
\end{eqnarray}
and $\zeta$ is a dimensionless parameter accounting for the fine structure splitting.

When neglecting the hyperfine interaction, the two states of the clock transition with $J = 0$ do not exhibit vector or tensor polarizabilities, as can be seen by setting $F = 0$ in equation~(\ref{eq:kj}). However, the hyperfine splitting of the excited states $\ket{\phi'}$:
\begin{equation}
	\nu_\hfs = \frac 1 2 A C + B\frac{\frac 3 4 C(C+1) - I(I+1)J'(J'+1)}{2I(2I - 1)J'(2J' - 1)},
\end{equation}
where $C = F'(F'+1) - I(I+1) - J'(J'+1)$ and $A$ and $B$ are magnetic dipole and electric quadrupole constants, gives rise to vector and tensor components. The relative magnitude of these latter components is expected to be on the order of the ratio between the hyperfine splitting $\nu_\hfs$ and the frequency of the optical transitions $\ket{\phi} \rightarrow \ket{\phi'}$~\cite{katori_ultrastable_2003}:
\begin{equation}
	\label{eq:approx}
	\frac{\alpha_{v,t}}{\alpha_s} \approx \frac{\nu_\hfs}{\nu_\text{optical}}.
\end{equation}
However, for a state $\ket{\phi}$ with $J = 0$, we can show that for any value of $I$, and with $J'=1$ according to the electric dipole selection rules:
\begin{equation}
	\sum_{F' = I - J'}^{I+J'} k_2C = 0 \qquad (J = 0,\ J' = 1),
\end{equation}
such that the contribution of the magnetic dipole term $A$ to the tensor polarizability is cancelled at first order in $\nu_\hfs$. Hence, the dominant contribution to the tensor shift is the quadrupole $B$, or the second order in $\nu_\hfs$, whichever is the largest. No such cancellation occurs for the vector shift. This property is particularly observed for $^{87}$Sr, as the excited states with the largest hyperfine splitting feature $B \ll A$, like $5s6s\ ^3S_1$ and $5s5p\ ^3P_1$. As a consequence, their contribution to the tensor polarizabilities of the clock states is orders of magnitude smaller than what could be expected from~(\ref{eq:approx}).

Using experimental measurements of the lifetimes of excited states, completed by theoretical estimates of the strength of atomic transitions~\cite{boyd_high_2007}, we can calculate the dynamic polarizabilities for various states of $^{87}$Sr at the magic wavelength ($\omega_l = 2\pi \times 368.6$~THz).

The scalar, vector and tensor polarizabilities of the fundamental state $5s^2\,^1S_0$ are derived by summing over the excited states $5s5p\,^3P_1$ and $5skp\,^1P_1$ with $k=5\ldots20$. They read, expressed in atomic units (a.u.):
\begin{equation}
\label{eq:alpha1S0}
\begin{array}{|c|c|c|}
	\hline \multicolumn{3}{|c|}{5s^2\,^1S_0 }          \\ \hline
	\alpha_s & \alpha_v           & \alpha_t           \\ \hline \hline
	279.8    & 4.75 \times 10^{-5} & 1.57 \times 10^{-5} \\ \hline
\end{array}
\end{equation}
The main contribution to the scalar polarizability comes from the $5s5p\,^1P_1$ state, while the vector and tensor terms are dominated by the $5s5p\,^1P_1$ and $5s5p\,^3P_1$ states. As expected, the ratio between the vector (resp. tensor) polarizability and the scalar polarizability is on the order of $10^{-7}$, comparable with the ratio between magnetic dipole term $A$ -- about 200~MHz (resp. the quadrupole term $B$ -- about 50~MHz) and optical frequencies -- about 500~THz.

\begin{figure}
	\begin{center}
		\includegraphics[width=0.95\columnwidth]{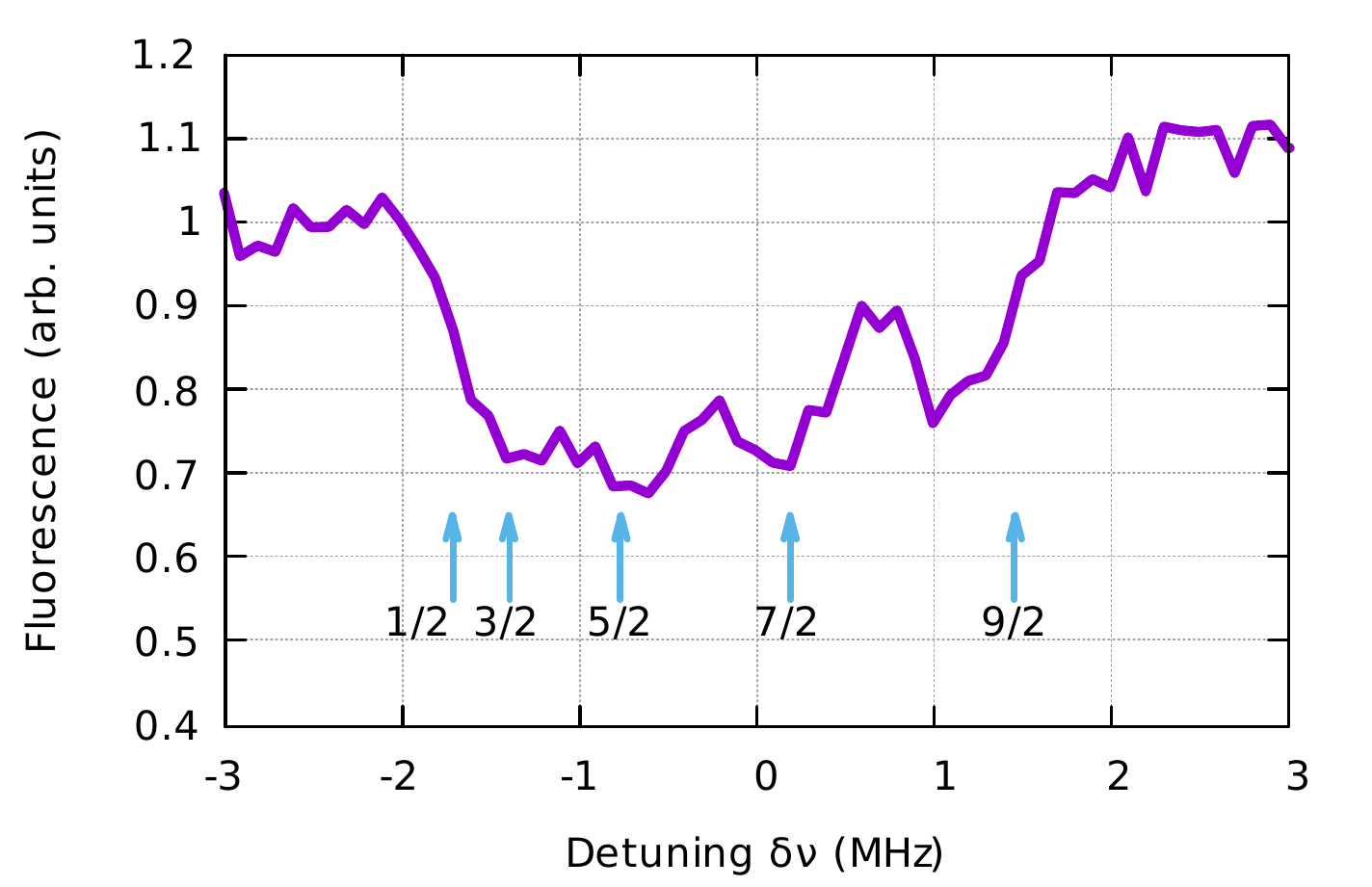}
	\end{center}
	\caption{\label{fig:3P1}Spectroscopy of the saturated $5s^2\,^1S_0\,\rightarrow\,5s5p\,^3P_1 (F=9/2)$ transition by depletion of $^{87}$Sr atoms in an optical lattice, for a linearly polarized trapping light, no magnetic field, and a trap depth of 1520 $E_R$, where $E_R = h\times3.47$~kHz is the recoil energy associated with the absorption of a lattice photon with frequency $\omega_l$. In this configuration, the polarizability operator~(\ref{eq:polemfact}) is already diagonal when choosing the quantization axis along the lattice polarization. Its eigen-values are $\alpha(m) = \alpha_s + \alpha_t \frac{3m^2 - F(F+1)}{F(2F-1)}$. The graph is centred on the light shift free frequency of the transition, and displays the splitting between magnetic sub-levels induced by the scalar and tensor polarizabilities of both states. The arrows indicate the theoretical light shift $\delta\nu = -\frac 1 4 \alpha(m) |\mathcal{E}|^2$ for each value of $|m|$, as calculated from equations~(\ref{eq:alpha1S0}) and~(\ref{eq:alpha3P1}).}
\end{figure}

The polarizability of the $5s5p\,^3P_1 (F=9/2)$ is estimated by summing over the $5sks\,^3S_1$ states with $k=6\ldots10$, the $5p^2\,^3P$ states and the $5skd\,^3D$ states with $k=4\ldots9$:
\begin{equation}
\label{eq:alpha3P1}
\begin{array}{|c|c|c|}
	\hline \multicolumn{3}{|c|}{5s5p\,^3P_1 (F=9/2)} \\ \hline
	\alpha_s & \alpha_v & \alpha_t \\ \hline \hline
	304.4    & -50.3   & -101.2    \\ \hline
\end{array}
\end{equation}
As expected from the fact that $J \neq 0$ for $^3P_1$, the vector and tensor polarizabilities are on the same order of magnitude as the scalar polarizability. Figure~\ref{fig:3P1} confronts these numerical values to the experimental spectroscopy of the $5s^2\,^1S_0\,\rightarrow\,5s5p\,^3P_1 (F=9/2)$ inter-combination line.

In the case of the excited clock state $5s5p\ ^3P_0$, a direct calculation is not sufficient because state mixing with the $5s5p\ ^3P_1 (F=9/2)$, $5s5p\ ^3P_2 (F=9/2)$ and $5s5p\ ^1P_1 (F=9/2)$ states alters its physical properties. Since these states have $J \neq 0$, their vector and tensor polarizabilities are large and may contribute to the polarizabilities of the $^3P_0$ state. This state mixing is written:
\begin{equation}
	\ket{\phi} = \sum_{p} c_p \ket{\phi_p}_0,
\end{equation}
where the coefficients $c_p$ are~\cite{boyd_nuclear_2007}:
\begin{equation}
\label{eq:mixingcoeff}
\begin{tabular}{|c||c|c|c|c|}
	\hline
	$\ket{\phi_p}_0$ & $5s5p\ ^3P_0$ & $5s5p\ ^3P_1$ & $5s5p\ ^3P_2$ & $5s5p\ ^1P_1$  \\
	& & $(F = 9/2)$ & $(F = 9/2)$ & $(F = 9/2)$ \\ \hline
	$c_p$ & 1. & $2.3 \times 10^{-4}$ & $ -1.38 \times 10^{-6}$ & $ -4.1 \times 10^{-6}$ \\ \hline
\end{tabular}.
\end{equation}

When state mixing is involved, expanding the matrix elements of the polarizability operator~(\ref{eq:pol}) is more involving, but the result is that equations~(\ref{eq:polemfact}) through~(\ref{eq:Bhfs}) remain valid expressions of these matrix elements, provided they are averaged over all $(\ket{\phi_p}_0, \ket{\phi_q}_0)$ combinations with a weight $c_p c_q^*$, and provided the branching ratios are replaced by:
\begin{eqnarray}
	\nonumber
	\bar{B}_\fs & = &  (-1)^{S_p + S_q + 2L' + J_p + J_q}\sqrt{(2J_p + 1)(2J_q + 1)}\\
	            & \times & (2 L'+1)\SixJSymbol{J',1,J_p}{L,S,L'}\SixJSymbol{J',1,J_q}{L,S,L'},
\end{eqnarray}
\begin{equation}
	\bar{B}_\hfs = (2F+1)(2J'+1)\SixJSymbol{F',1,F}{J_p,I,J'}\SixJSymbol{F',1,F}{J_q,I,J'}.
\end{equation}
Using these expressions, we can show that for $J_p = 0$ and $J_q = 1$, $\sum_{F' = |I - J'|}^{I+J'} k_2 = 0$ such that this combination has a negligible contribution to the tensor shift, and for $J_p = 0$ and $J_q = 2$, $\sum_{F' = |I - J'|}^{I+J'} k_1 = 0$ such that this second combination has a negligible contribution to the vector shift.

Among all the $(\ket{\phi_p}_0, \ket{\phi_q}_0)$ combinations involving the states listed in the table of equation~(\ref{eq:mixingcoeff}), only a few lead to a non-negligible contribution. They are listed in bold characters in the following table:
\begin{equation}
\label{eq:alpha3P0}
\begin{array}{|c|c|c|c|c|}
	\hline \multicolumn{5}{|c|}{5s5p\,^3P_0}                                  \\ \hline
	\ket{\phi_p}_0 & \ket{\phi_q}_0 & \alpha_s            &  \alpha_v            &  \alpha_t            \\ \hline \hline
	^3P_0 & ^3P_0 & \mathbf{288.8}      &  \mathbf{9.3 \times 10^{-3}} &  1.9 \times 10^{-6} \\ \hline
	^3P_1 & ^3P_1 & 2 \times 10^{-5} & 2.7 \times 10^{-6} & \mathbf{ -5.4 \times 10^{-6}} \\ \hline
	^3P_0 & ^3P_1 & 2 \times 10^{-6} & \mathbf{ 1.82 \times 10^{-1}} & -1.5 \times 10^{-6} \\ \hline
	^3P_0 & ^3P_2 & \leq  10^{-13}      & -1.8 \times 10^{-8} &  \mathbf{3.70 \times 10^{-4}} \\ \hline \hline
	\multicolumn{2}{|r|}{\textrm{Total}}
          & 288.8               &  1.91 \times 10^{-1} &  3.65 \times 10^{-4} \\\hline
\end{array}
\end{equation}
In this table, the $(\ket{\phi_p}_0, \ket{\phi_q}_0)$ and $(\ket{\phi_q}_0, \ket{\phi_p}_0)$ contributions for $p \neq q$ are grouped together. As expected from the theoretical considerations above, the tensor term of the $(^3P_0,^3P_0)$ configuration is negligible, and the main contribution to the vector and tensor polarizabilities respectively come from the mixing combinations $(^3P_0,^3P_1)$ and $(^3P_0,^3P_2)$.

To express these polarizabilities in units more adapted to experimental purposes, we note $U_0 = \frac 1 4 \alpha_s|\mathcal{E}|^2$ the trap depth. At the magic wavelength, $\alpha_s$ and hence $U_0$ are by definition identical for the two clock states. The numerical values of $\alpha_s$ shown in equations~(\ref{eq:alpha1S0}) and~(\ref{eq:alpha3P0}) show that this fact is approximately rendered by our calculation, which also gives an order of magnitude of a few percent for its accuracy, arising from uncertainties on the decay rates from excited states. For the vector and tensor polarizabilities, we introduce the coefficients:
\begin{equation}
		\kappa^v = -\frac{\alpha_v}{\alpha_s} \frac{1}{h} \frac{1}{2F} \quad \text{and} \quad \kappa^t = -\frac{\alpha_t}{\alpha_s} \frac{1}{h} \frac{1}{2F(2F-1)}.
\end{equation}

The following table gathers the theoretical values for $\kappa^v$ and $\kappa^t$ for the two clock states, as well as their difference $\Delta\kappa^{v,t} = \kappa^{v,t}(^3P_0) - \kappa^{v,t}(^1S_0)$:
\begin{equation}
\label{eq:Dkappath}
\begin{array}{|c|c|c|}
	\hline
	\ket{\phi}&\kappa^v\ (\text{mHz}/E_R)& \kappa^t \ (\mu\text{Hz}/E_R)\\
	\hline\hline
	^3P_0 & -255 & -60.9 \\
	\hline
	^1S_0 & -6.5 \times 10^{-2} & -2.7\\
	\hline\hline
	\mathbf{^3P_0 - {^1S_0}} & \mathbf{-255} & \mathbf{-58.2} \\
	\hline
\end{array}
\end{equation}

\section{Experimental determination of the polarizabilities}

\begin{figure}
	\includegraphics[width=\columnwidth]{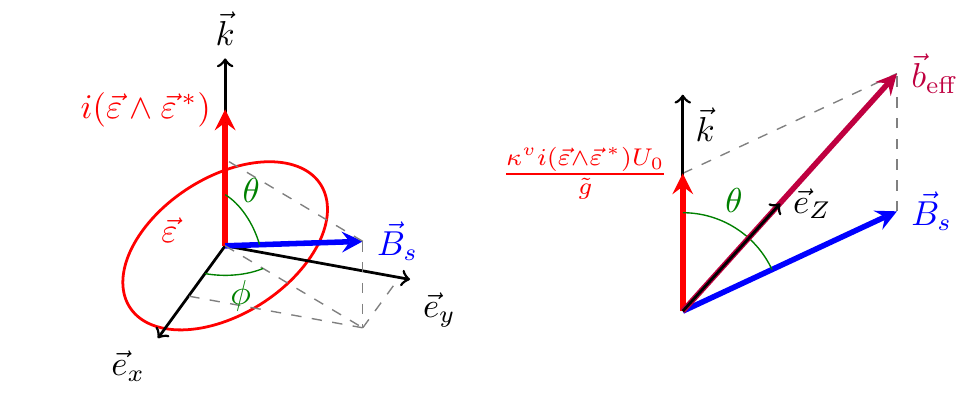}
	\caption{\label{fig:polmag}Left: parametrization of the light polarization $\e$ and the static bias magnetic field $\vec B_s$. $\vec e_x$ (resp. $\vec e_y$) is a unit vector along the major (resp. minor) axis of the trapping light polarization $\e$, such that the polarization ellipsis (red) is included in the $(\vec e_x,\vec e_y)$ plane. Right: projection in the plane defined by the trapping light wave vector $\vec k$ and $\vec B_s$. The fictitious magnetic field representing the vector light shift adds to $\vec B_s$ to form an effective magnetic field $\beff$. If the vector component of the Hamiltonian dominates its tensor component, the Hamiltonian is diagonal for a quantization axis $\vec e_Z$ defined along $\beff$. When the polarization $\e$ is elliptical, the quantization axis therefore depends on the trap depth $U_0$, introducing non-linear effects in the Zeeman shift and average clock frequency as a function of $U_0$.}
\end{figure}
During the clock operation, a static bias magnetic field is applied to split the magnetic sub-levels. The energy levels are then the eigenvalues of the full Hamiltonian written in equation~(\ref{eq:H}). Because $\hat {\vec F}$ is a vector operator, the Zeeman Hamiltonian adds up to the vector component of the polarizability operator. Therefore, the vector component of the Hamiltonian is equivalent to the action of an effective magnetic $\beff$ field equal to the vector sum of the static magnetic field $\vec B_s$ and the fictitious magnetic field that represents the vector light shift (see figure~\ref{fig:polmag}):
\begin{equation}
	\label{eq:beff}
	\beff = \frac{\kappa^v \iee U_0}{\tg} +  \vec B_s,
\end{equation}
where $\tg = g\mu_B/h$. If the quantization axis $\vec e_Z$ is chosen along the effective magnetic field (\emph{i.e} $\vec e_Z = \beff/||\beff||$), the vector component of the Hamiltonian is diagonal in the basis $\left\{\ket{m}, -F \leq m \leq F\right\}$ of eigen-vectors of $F_Z$. However, the tensor part of the Hamiltonian is \emph{a priori} not diagonal in this basis. Yet, since $\kappa^v \gg \kappa^t$, if we assume that the bias field is large enough for the Zeeman splitting to be much larger than the energy splitting due to the tensor shift (For a trap depth $U_0 \sim 100\, E_R$, this assumption is satisfied if $B_s \gg 1~\mu$T), the states $\left\{\ket{m}\right\}$ remain approximate eigen-states of the full Hamiltonian. Expanding the tensor product in equation~(\ref{eq:polemfact}) then yields the energy shift $\delta \nu$ of level $\ket{m}$ of a given hyperfine manifold:
\begin{equation}
	\label{eq:numarbB}
	\delta\nu(m) = -\frac{U_0}{h} +  m\tg||\beff|| + \kappa^t \beta U_0,
\end{equation}
with
\begin{equation}
	\beta = (3|\ecez|^2-1) (3m^2 - F (F+1)).
\end{equation}

The middle term of equation~(\ref{eq:numarbB}) includes both the vector light shift and the Zeeman shift. It can be isolated from the scalar and tensor terms by measuring the half difference $Z_s(m) = \frac 1 2 \left(\delta\nu(m) - \delta\nu(-m)\right)$ between opposite magnetic sub-levels:
\begin{equation}
	\label{eq:nlzs}
	Z_s = m\tg||\beff|| = m\sqrt{\left(\tg B_s\cos\theta + \kappa^v\xi U_0\right)^2 + \left(\tg B_s\sin\theta\right)^2},
\end{equation}
where $\xi = ||\iee||$ is the degree of circular polarization of the trapping light, and $\theta$ is the angle between the wave vector $\vec k$ and the bias field $\vec B_s$ (as shown on figure~\ref{fig:polmag}). The Taylor expansion of the later equation to second order in $U_0$ yields:
\begin{equation}
	\label{eq:nlzstaylor}
	Z_s \simeq m\tg B_s + m\kappa^v\xi \cos\theta\,U_0 + m\sin^2\theta \frac{(\kappa^v\xi)^2}{2\tg B_s}\,U_0^2.
\end{equation}
To experimentally evaluate the differential vector polarizability coefficient $\Delta\kappa^v$, we measured $\Delta Z_s = Z_s(^3P_0) - Z_s(^1S_0)$ for different trapping depths with $m = \pm 9/2$ and a circular polarization for the trapping light ($|\xi| = 1$). From these data we extrapolated the derivative of the Zeeman shift at zero trap depth $\partial \Delta Z_s/ \partial U_0 (U_0 = 0) = m\Delta\kappa^v\xi \cos\theta$. The trap depth is measured by observing the longitudinal motional side-bands of the trapped atoms~\cite{PhysRevA.80.052703}: their spacing and shape give a direct measurement of the trap depth (at the trap centre) and of longitudinal and transverse temperatures of the atoms~\cite{letargat:pastel-00553253}. From these three parameters, we can deduce the average trap depth experienced by the atoms.

Repeating the experiment for various values of $\theta$ enables to deduce the differential coefficient, as reported in~\cite{westergaard2011lattice}:
\begin{equation}
	|\Delta \kappa^v| = 0.22 \pm 0.05 \text{ Hz}/E_R,
\end{equation}
in agreement with the theoretical estimate reported in equation~(\ref{eq:Dkappath}). The uncertainty is limited by the knowledge of the polarization state $\e$ in the lattice cavity. In the usual clock operation, $\vec B_s$ is orthogonal to the wave vector $\vec k$ (\emph{i.e.} $\cos\theta \simeq 0$). This configuration minimizes the first order term of equation~(\ref{eq:nlzstaylor}), but it also maximizes its quadratic term. This latter term can therefore easily be observed, even for moderate trapping depths below 200~$E_R$.
\begin{figure}
\begin{center}
	\includegraphics[width=\columnwidth]{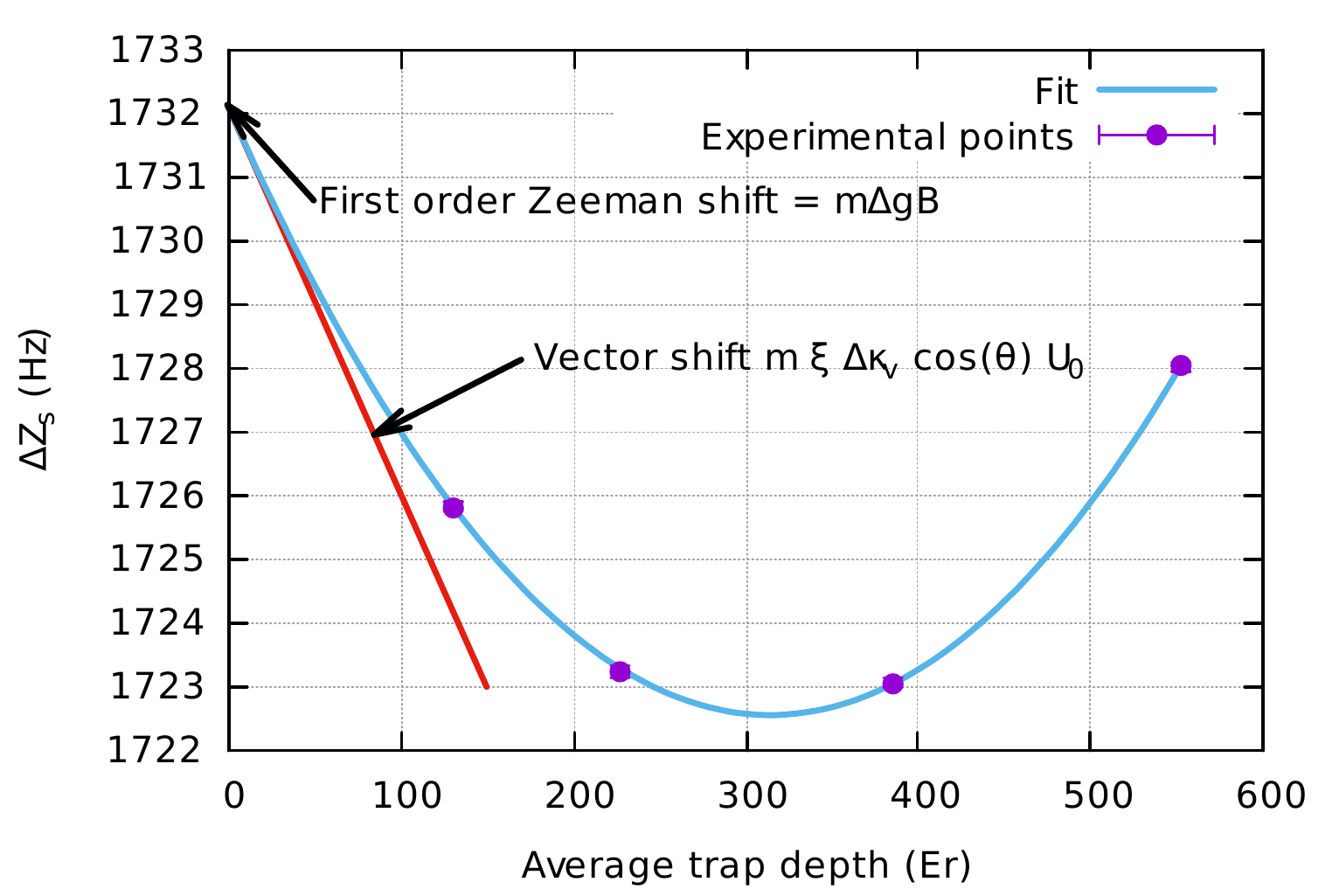} 
	\caption{\label{fig:vector}Differential Zeeman shift $\Delta Z_s$ of the $^ {87}$Sr clock resonance as a function of the trap depth $U_0$ for an elliptical polarization $|\xi| \simeq 0.9$ and $|m = 9/2|$. The experimental data points are fit with equation~(\ref{eq:nlzs}), assuming $\kappa^v(^1S_0) = 0$. These four data points are sufficient to determine the three free parameters of this equation that describe the geometry of the experimental setup, here $B_s = 355.0 \pm 0.2 ~\mu\text{T}$, $\xi\kappa^v(^3P_0) = 212 \pm 2~\text{mHz}/E_R$, and $\theta = \text 63.9 \pm 0.5~\text{mrad}$.}
\end{center}
\end{figure}
\begin{figure}
\begin{center}
	\includegraphics[width=\columnwidth]{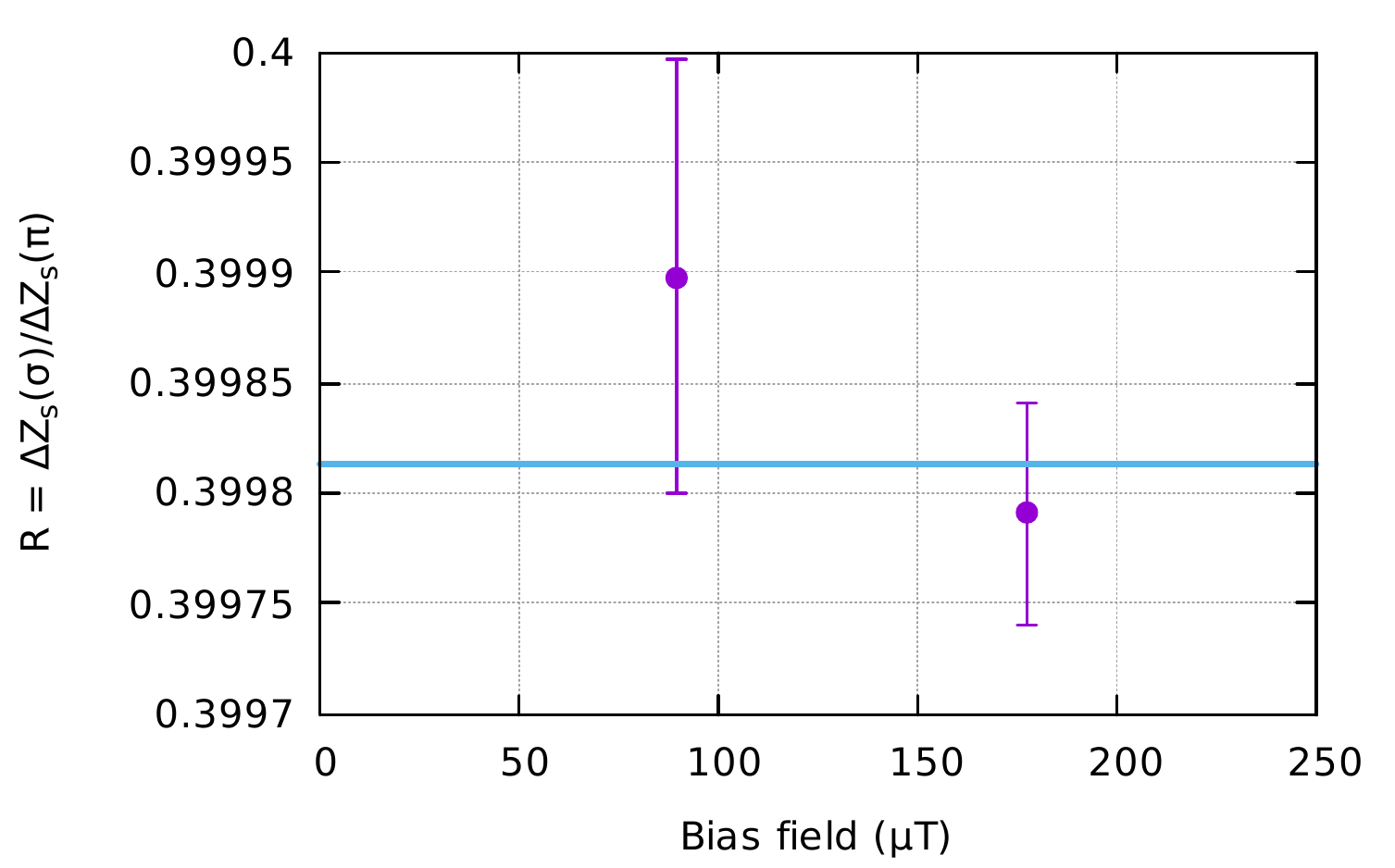}
	\caption{\label{fig:lande}Ratio $R$ between the differential Zeeman shifts $\Delta Z_s$ (for linear polarization, so that no dependence in $U_0$ is observed) for $\sigma (|m| = 9/2 \rightarrow |m| = 7/2)$ and $\pi (|m| = 9/2 \rightarrow |m| = 9/2)$ transitions. This ratio is related to the ratio between Land\'e factors by $\tg(^3P_0)/\tg(^1S_0) = 9(R-1)/(9R-7)$.}
\end{center}
\end{figure}
Figure~\ref{fig:vector} shows such a quadratic dependence for an elliptical polarization. In order to fit the data points of this figure, the two Land\'e factors of $^3P_0$ and $^1S_0$ have to be known. We performed a precise measurement of the ratio between these coefficients by locking the clock on $\sigma^\pm$ and $\pi$ transitions, as shown on figure~\ref{fig:lande}:
\begin{equation}
	\frac{\tg(^3P_0)}{\tg(^1S_0)} = 1.58794 (7).
\end{equation}
Given the Land\'e factor of the fundamental state $\tg(^1S_0) = 184.4 (1)$~\cite{muI,boyd_nuclear_2007}, this gives:
\begin{equation}
	\tg(^3P_0) = 2.928 (2) \text{MHz/T}, \text{ and } \Delta\tg = 1.0842 (7) \text{MHz/T}.
\end{equation}
in agreement with previously reported values~\cite{boyd_nuclear_2007}.

The measurement of the tensor shift was conducted with a purely linear polarization for the trapping light, and therefore with a quantization axis along the bias field $\vec B_s$. The tensor component is extracted by measuring the average clock frequency shift for various values of $\beta$. The results are reported in table~\ref{tab:beta}:
\begin{eqnarray}
	\nonumber
	\Delta \nu & = & \frac{\nu(^3P_0,m) - \nu(^1S_0,m) + \nu(^3P_0,-m) - \nu(^1S_0,-m)}{2} \\ & = & (\Delta \kappa^s + \beta \Delta\kappa^t)U_0.
	\label{eq:Deltanu}
\end{eqnarray}
\begin{table*}
	\begin{center}
		\begin{tabular}{|r||c|c|c|c|c|c|c|c|c|c|c|c|c|c|c|}
			\hline
			Meas. \# & 1 & 2 & 3 & 4 & 5 & 6 & 7 & 8 & 9 & 10 & 11 & 12 & 13 \\ \hline
			\hline
			$\sin \theta$ &
			1             &
			1             &
			0.92          &
			0.92          &
			1             &
			1             &
			1             &
			1             &
			0.99          &
			0.99          &
			0.955         &
			0.955         &
			1             \\ \hline
			$\cos\phi$ &
			0.602      &
			0.602      &
			0.602      &
			0.602      &
			0          &
			0          &
			1          &
			1          &
			1          &
			1          &
			1          &
			1          &
			0          \\ \hline
			$m$   &
			9/2   &
			7/2   &
			9/2   &
			7/2   &
			9/2   &
			7/2   &
			9/2   &
			7/2   &
			9/2   &
			7/2   &
			9/2   &
			7/2   &
			9/2   \\ \hline
			$\beta$  &
			3.12     &
			1.04     &
			-2.96    &
			-0.99    &
			-36      &
			-12      &
			72       &
			24       &
			70.0     &
			23.3     &
			63.05    &
			21.0     &
			-36      \\ \hline
		\end{tabular}
	\end{center}
	\caption{\label{tab:beta}Geometrical configurations chosen to measure the tensor shift. A strictly linear polarization is selected by the enhancement cavity of the optical lattice by using an intra-cavity birefringent element. For this polarization, the $\beta$ coefficient reduces to $\beta = (3\sin^2 \theta \cos^2\phi - 1) (3m^2 - F(F+1)) = (3\cos^2 - 1) \alpha (3m^2 - F(F+1))$ where $\alpha$ is the angle between the polarization and the bias field $\vec B_s$. We switch between $\cos\phi = 1$ and $\cos\phi = 0$ by changing the polarization eigen-mode of the lattice cavity. The angles are derived from the relative values of the Zeeman shift when the magnetic field is changed.}
\end{table*}
where:
\begin{equation}
	\Delta \kappa^s = -\frac{\Delta \alpha_s}{\alpha_s} = -\frac{\alpha_s(^3P_0) - \alpha_s(^1S_0)}{\alpha_s}
\end{equation}
is the remaining differential scalar light shift due to a possible detuning of the trapping light from the magic wavelength. The resulting value for the tensor shift coefficient is then, as reported in~\cite{westergaard2011lattice}:
\begin{equation}
	\Delta \kappa^t_\textrm{exp} = \frac{\partial \Delta\nu}{\partial\beta} =  (-57.7\pm 2.3) \, \mu\textrm{Hz}/E_r,
\end{equation}
in agreement with the theoretical value shown in equation~(\ref{eq:Dkappath}) well within the experimental error bar.

We now consider the behaviour of the tensor light shift when the light polarization $\e$ is not perfectly linear. In this case, the quantization axis $\vec e_Z$ for the excited state $^3P_0$ is aligned with the effective magnetic field $\beff$ and thus varies with the trap depth $U_0$. Consequently, the $\beta$ coefficient depends on $U_0$ through:
\begin{equation}
	|\ecez|^2 = \left|\frac{\e \cdot \vec B_s}{||\beff||}\right|^2 \simeq |\eceB|^2\left(1 - 2\frac{\kappa^v\xi \cos\theta\,U_0}{\tg B_s} + o\left(U_0\right)\right).
\end{equation}
where $\vec e_{B_s}$ is the unit vector along $\vec B_s$. Because of this coupling effect between the vector and tensor light shifts, the average clock frequency contains a frequency shift $\Delta \nu_{vt}$ quadratic with the trapping potential reading:
\begin{equation}
	\Delta \nu_{vt} = -\gamma_{vt} |\eceB|^2 \xi\cos\theta\,U_0^2,
\end{equation}
with
\begin{equation}
 \gamma_{vt} = \left(3m^2 - F(F+1)\right)\frac{6\kappa^t(^3P_0)\kappa^v(^3P_0)}{\tg(^3P_0) B_s},
\end{equation}
assuming that $^1S_0$ exhibits no vector shift. For $^{87}$Sr we have $\gamma_{vt} = 11\ \mu\text{Hz}/E_R^2$ for $B_s = 0.1$~mT. This quadratic frequency shift on the average clock frequency is three orders of magnitude smaller than the quadratic term of the Zeeman shift written in equation~(\ref{eq:nlzstaylor}). But this effect is quite significant when compared to the hyper-polarizability coefficient $\gamma = 0.4\ \mu\text{Hz}/E_R^2$~\cite{westergaard2011lattice,eftf2012}. However, it is dramatically reduced when the polarization is linear, $\vec B_s\,\bot\,\vec k$ and $\vec B_s\,\bot\,\e$. Its experimental demonstration is for now challenging, as it is predominant in a configuration where the vector light shift is large and blurs the atomic resonances.

Finally, repeated measurements of the total light shift over two years with our two strontium clocks~\cite{le_targat_experimental_2013} confirmed the experimental value for the magic wavelength published in~\cite{westergaard2011lattice}, for which the differential scalar light shift is cancelled (\emph{i.e.} $\Delta \kappa^s = 0$):
\begin{equation}
	\nu_\text{magic} = 368\,554\,725 \pm 3 \text{ MHz}.
\end{equation}
The uncertainty on this value is limited by our knowledge of $\beta$ used to subtract the tensor light shift from the total light shift from equation~(\ref{eq:Deltanu}). In addition, we measured the sensitivity of the scalar light shift with the trapping light frequency $\nu_\text{latt}$:
\begin{equation}
	\frac{\partial \Delta \kappa^s}{\partial \nu_\text{latt}} = -15.5 \pm 1.1 \ \mu\text{Hz}/E_R/\text{MHz}.
\end{equation}
This value is in agreement with the estimation given in~\cite{PhysRevLett.109.263004} from Monte-Carlo simulation using transition strengths and scalar polarizabilities of the clock states at various frequencies.

\section*{Conclusion}

In this paper, we have conducted a theoretical estimate of the vector and tensor polarizabilities of the $^{87}$Sr clock states in agreement with experimental measurements. We have shown that the differential vector polarizability is largely due to state mixing of the excited clock state $^3P_0$ with $^3P_1$. The tensor polarizability is mainly due to state mixing of  $^3P_0$ with $^3P_2$, with a small contribution from the hyperfine structure of $^1P_1$ on $^1S_0$. We also described a non-linear behaviour of these light shifts that are largely cancelled in normal clock operation, but that have to be considered as the accuracy of optical lattice clocks continues to improve.

SYRTE is UMR CNRS 8630 between Centre National de la
Recherche Scientifique, Universit\'e Pierre et Marie Curie, and
Observatoire de Paris. The Laboratoire National de M\'etrologie
et d'Essais is the French National Metrology Institute. This work is supported by CNES, IFRAF and Nano-K-Conseil R\'egional \^Ile-de-France and ESA under the SOC project.

\bibliographystyle{unsrt}
\bibliography{polar}

\end{document}